\documentclass[conference,draftcls,onecolumn]{IEEEtran}	

\usepackage[latin1]{inputenc}
\usepackage[T1]{fontenc}
\usepackage[english]{babel}
\usepackage{amssymb,amsmath,amsfonts}
\usepackage{times}
\usepackage{graphicx}
\usepackage{color}
\usepackage{verbatim}

\usepackage{pgf}		
\usepackage{tikz}
\usetikzlibrary{decorations}

\usepackage{stmaryrd}

\newtheorem{defi}	{Definition}
\newtheorem{theo}	{Theorem}
\newtheorem{prop}	{Proposition}
\newtheorem{coro}	{Corollary}

\newcommand{\cA}{{\mathcal A}}
\newcommand{\cB}{{\mathcal B}}
\newcommand{\cE}{{\mathcal E}}
\newcommand{\cR}{{\mathcal R}}
\newcommand{\cU}{{\mathcal U}}
\newcommand{\cV}{{\mathcal V}}

\newcommand{\bN}{{\mathbb N}}
\newcommand{\bR}{{\mathbb R}}
\newcommand{\bE}{{\mathbb E}}

\newcommand{\pr}[1]{\operatorname{Pr}\left\{#1\right\}}

\newcommand{\mkv}{-\!\!\!\!\minuso\!\!\!\!-}
\newcommand{\typ}[1]{T_\epsilon^n(#1)}

\providecommand{\norm}[1]{\lVert#1\rVert}
\IEEEoverridecommandlockouts

\begin{document}

\title{Secure Lossy Source Coding\\ with Side Information at the Decoders} 

\author{
\IEEEauthorblockN{Joffrey Villard and Pablo Piantanida}
\IEEEauthorblockA{Department of Telecommunications, SUPELEC\\
91192 Gif-sur-Yvette, France\\
Email: \{joffrey.villard,pablo.piantanida\}@supelec.fr}
\thanks{The work of J. Villard is supported by DGA (French Armement Procurement Agency). This research is partially supported by the FP7 Network of Excellence in Wireless COMmunications NEWCOM++.}
}    

\date{September 2010} 

\maketitle

\renewcommand{\leftmark}{\MakeUppercase{to be presented at Allerton 2010}}	
\renewcommand{\rightmark}{} 												

\begin{abstract}
This paper investigates the problem of  secure lossy source coding in the presence of an eavesdropper with arbitrary correlated side informations at the legitimate decoder (referred to as Bob) and the eavesdropper (referred to as Eve). This scenario consists of an encoder that wishes to compress a source to satisfy the desired requirements on: (i) the distortion level at Bob and (ii) the equivocation rate at Eve. It is assumed that the decoders have access to correlated sources as side information. For instance, this problem can be seen as a generalization of the well-known Wyner-Ziv problem taking into account the security requirements. A complete characterization of the rate-distortion-equivocation region for the case of arbitrary correlated side informations at the decoders is derived. Several special cases of interest and an application example to secure lossy source coding of binary sources in the presence of binary and ternary side informations are also  considered. It is shown that the statistical differences between the side information at the decoders and the presence of non-zero distortion at the legitimate decoder can be useful in terms of secrecy. Applications of these results arise in a variety of distributed sensor network scenarios.
\end{abstract}

\section{Introduction}

Consider the problem of compressing correlated sources at sensor nodes 
in a distributed fashion where the sensors may wish to communicate 
with each other on a wireless network. Assume also that each of these sensors 
can have access to a correlated observation to the source or random field of interest. 
This observation can be used as side information available at the decoder to minimize 
the distortion between the original source and the estimate at the legitimate decoder (referred to as Bob).
In addition to this, we assume that each of the encoders (referred to as Alice) wishes to leak the least possible 
amount of information about its source to an eavesdropper (referred to as Eve), e.g. 
an untrusted sensor, who may capture such information during the communication between nodes. 

The above scenario involves many of the major information-theoretic issues on source and channel coding problems. 
In terms of source coding, Slepian-Wolf~\cite{slepian1973noiseless} and  Wyner-Ziv~\cite{wyner1976rate}  introduced the problem of source coding with side information at the decoder. This topic has been the focus of intense study and some remarkable progress has already been made in theoretical and practical aspects. On the other hand,  extensive research has been done  during the recent years on secure communications over noisy channels. The traditional focus was on cryptography, based on computational complexity where  security only depends on the intractability assumption of NP-complete problems that must be solved prior to  decoding. Another approach is the information-theoretic notion of secrecy, introduced by Shannon in~\cite{shannon1949communication}, where security is measured  through the equivocation rate (i.e. the remaining uncertainty about the message) at the eavesdropper. The wiretap channel was introduced by Wyner~\cite{wyner1975wire}, who showed that it is possible to send information at a positive rate with perfect secrecy  as long as the channel of the legitimate user is less noisy than the channel of the   eavesdropper.  Csisz\`ar-Korner~\cite{csiszar1978broadcast} extends this result to the setting of  general broadcast channels with any arbitrary equivocation rate. Several extensions of the wiretap and fading channels have been done (cf. \cite{it2008special,liang2009information}  and references therein). So far, very few work has been reported on source coding (or compression) problems with security constraints.

One can identify two approaches in the literature on secure source coding. In fact, it is assumed either that there already exists a secure rate-limited channel between Alice and Bob, which allows the system to use secret keys, or the decoders have access to side information about the source. In the scenario of secret key sharing,  both lossless and lossy compression have been studied in various contexts~\cite{yamamoto1983source,yamamoto1988rate,yamamoto1994coding,yamamoto1997rate,liu2009securing,merhav2006shannon}.
For the second scenario  where side information is available at both decoders, the case of lossless source coding  has been recently studied in~\cite{prabhakaran2007secure,gunduz2008secure,gunduz2008lossless,tandon2009securea}. Whereas the general lossy source coding problem has not been fully solved, some particular cases can be derived as part of previous work. It is important to mention here that if the side informations between Bob and Eve are degraded then the result follows as a special case of~\cite{merhav2008shannon}.

In this paper, we investigate the problem of secure lossy source coding of memoryless sources in the presence of an eavesdropper  with different correlated side informations at the decoders of Bob and Eve, as it is shown in Fig.~\ref{fig:schema}.  In this setting  the channels between encoder and decoders are assumed to be noiseless so that they cannot provide any advantage to increase security. Our goal is to understand the minimum amount of information that needs to be revealed to Eve to satisfy the distortion constraint at Bob. We provide a complete characterization of the rate-distortion-equivocation region for the case of arbitrary correlated side informations. Several special cases of interest are also considered. As an application example, we consider the case of secure lossy source coding of a binary source, where the side information at Bob (resp. Eve) is the output of a binary erasure channel (resp. a binary symmetric channel) with the source as the input. This model is of interest since neither Bob nor Eve can always be a lessnoisy decoder.  

The organization of this paper is as follows. Section II states definitions along with the main results, while Section III provides several special cases and discussion. The sketch of the proofs are relegated to Section IV. Finally, Section V presents an application example to binary sources and  Section VI summarizes the paper.

\subsection*{Notations}
For any sequence~$(x_i)_{i\in\bN^*}$, notation $x_k^n$
stands for the collection $(x_k,x_{k+1},\dots, x_n)$.
$x_1^n$ is simply denoted by $x^n$.
By extension, for any subset $J\subset\{1,\dots,n\}$,
notation $x_J$ stands for the collection $(x_j)_{j\in J}$.
The cardinality of an alphabet is denoted by $\norm{\cdot}$. 
For every $\epsilon>0$, we denote $\epsilon$-typical and conditional $\epsilon$-typical sets by $\typ{X}$ and $\typ{Y|x^n}$, respectively. 
Following~\cite{cover2006elements}, entropy is denoted by $H(\cdot)$ and mutual information by $I(\cdot;\cdot)$.
Let $X$, $Y$ and $Z$ be three random variables on some alphabets with probability distribution~$p$.
If $p(x|y,z)=p(x|y)$ for each $x,y,z$, then they form a Markov chain, which is denoted by $X\mkv Y\mkv Z$.
For each $x\in\bR$, notation $[x]_+$ stands for $\max(0;x)$.
For each $a,b\in[0,1]$, $a\star b = a(1-b) + (1-a)b$.

\section{Problem Definition and Main Results}

\begin{figure}[!b]
\centering
\begin{tikzpicture}
	\node	(A) 		at (0,0) 					 		{$A^n$};
	\node	(alice)		at (1.5,0)	[rectangle,draw,blue] 	{Alice};
	\node	(B) 		at (6,2) 					 		{$B^n$};
	\node	(bob) 		at (6,1)	[rectangle,draw,blue] 	{Bob};
	\node	(hatA) 		at (7.5,1)	[right,blue]	 		{$\hat A^n\,:\ d(A^n,\hat A^n)\lesssim D$};
	\node	(eve) 		at (6,-1)	[rectangle,draw,red]	{Eve};
	\node	(E) 		at (6,-2)					 		{$E^n$};
	\node	(D) 		at (7.5,-1)	[right,red]				{$\frac1n H(A^n|W E^n)\gtrsim\Delta$};
		
	\draw[->]	(A)			to (alice);
	\draw[->]	(B)			to (bob);
	\draw[->]	(alice)		to node[anchor=south]{$W$} node[anchor=north]{(rate $R$)} (5,0) to (5,1) to (bob);
	\draw[->,blue]	(bob)	to (hatA);
	\draw[->]	(5,0)		to (5,-1)to (eve);
	\draw[->]	(E)			to (eve);
	\draw[->,red]	(eve)	to (D);
\end{tikzpicture}
\caption{Secure lossy source coding in the presence of side information at the decoders.}
\label{fig:schema}
\end{figure}
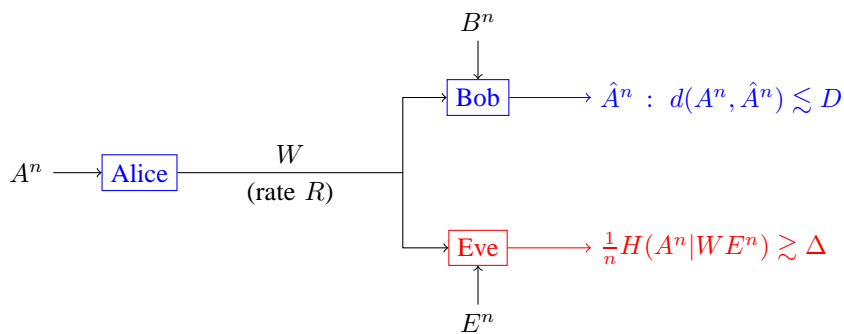

\subsection{Problem Definition}
In this section, we give a more rigorous formulation of the context depicted in Fig.~\ref{fig:schema}.
Let $\cA$, $\cB$ and $\cE$ be three finite sets. 
Alice, Bob and Eve observe the sequences of random variables 
$(A_i)_{i\in\bN^*}$, $(B_i)_{i\in\bN^*}$ and $(E_i)_{i\in\bN^*}$
respectively, which take values on $\cA$, $\cB$ and $\cE$, resp.
For each $i\in\bN^*$, the random variables $A_i$, $B_i$ and $E_i$
are distributed according to the joint distribution $p(a,b,e)$ on
$\cA\times\cB\times\cE$.
Moreover, they are independent across time $i$.

Let $d : \cA\times\cA \to [0\,;d_{max}]$ be a finite distortion measure
\emph{i.e.}, such that $0\leq d_{max} < \infty$.
We also denote by $d$ the component-wise mean distortion on $\cA^n\times\cA^n$
\emph{i.e.}, for each $a^n,b^n\in\cA^n$, $d(a^n,b^n) = \frac1n\,\sum_{i=1}^n d(a_i,b_i)$.

\begin{defi}
An $(n,R)$-code for source coding in this setup is defined by
\begin{itemize}
\item An encoding function at Alice $f : \cA^n \to \{1,\dots,2^{nR}\}$,
\item A decoding function at Bob $g : \{1,\dots,2^{nR}\}\times\cB^n \to \cA^n$.
\end{itemize}
\end{defi}

\begin{defi}
A tuple $(R,D,\Delta)\in\bR_+^3$ is said to be \emph{achievable} if,
for any $\epsilon>0$, there exists an $(n,R+\epsilon)$-code $(f,g)$ such that:
\begin{eqnarray*}
\bE\left[ d(A^n,g(f(A^n),B^n)) \right]	&\leq& D+\epsilon \ ,\\
\dfrac1n\,H(A^n|f(A^n),E^n) 			&\geq& \Delta-\epsilon \ .
\end{eqnarray*}
The set of all achievable tuples is denoted by $\cR^*$
and is referred to as the \emph{rate-distortion-equivocation region}.
\end{defi}

\subsection{Main Result}
The proof of the following theorem is given in Sections~\ref{sec:achievability}
and~\ref{sec:converse}.

\begin{theo}
\label{th:region}
Region $\cR^*$ is the set of all tuples $(R,D,\Delta)$ such that
there exist random variables $U$, $V$ on some finite sets $\cU$, $\cV$, respectively,
and a function $\hat A : \cV\times\cB \to \cA$
such that $U\mkv V\mkv A\mkv (B,E)$ form a Markov chain and
\begin{eqnarray*}
R 		&\geq& I(V;A|B) \ ,\\
D 		&\geq& \bE[d(A,\hat A(V,B))] \ ,\\
\Delta	&\leq& \Big[ H(A|VB) + I(A;B|U) - I(A;E|U) \Big]_+ \ .
\end{eqnarray*}
\end{theo}

The first two inequalities in Theorem~\ref{th:region} are classical
in rate-distortion theory.
Let us give some intuition on the third one.
The first term $H(A|VB)$ corresponds to the equivocation rate at Bob.
Alice thus exploits the available distortion at Bob to increase the equivocation rate at Eve.
Moreover, for a given random variable $V$, which determines the rate $R$ and the distortion level $D$,
auxiliary variable~$U$ may be tuned to make Bob \emph{more capable} than Eve \emph{i.e.},
maximize $I(A;B|U)-I(A;E|U)$.
This quantity represents the gain (or the loss) at Eve in terms of equivocation rate.

The following proposition gives upper bounds on the cardinalities of alphabets $\cU$ and $\cV$. The proof is omitted here and will be provided in an extended version of this paper.

\begin{prop}
\label{prop:card}
In the single-letter characterization of the rate-distortion-equivocation region~$\cR^*$
given by Theorem~\ref{th:region}, it suffices to consider sets~$\cU$ and~$\cV$ such that 
$\norm{\cU}\leq\norm{\cA}+2$ and $\norm{\cV}\leq(\norm{\cA}+2)(\norm{\cA}+1)$.
\end{prop}

\section{Special Cases of Interest}

In this section, we derive optimal regions of some special cases of Theorem~\ref{th:region}.

\subsection{Lossless Secure Source Coding}

The lossless secure source coding problem corresponds to a zero
distortion level at Bob ($D=0$). 
In this case, the following corollary, 
which can also be found in~\cite{prabhakaran2007secure,gunduz2008secure},
directly follows from Theorem~\ref{th:region}
(simply set $V=A$):

\begin{coro}
\label{th:lossless}
A tuple $(R,0,\Delta)$ is achievable i.f.f.
there exists a random variable $U$ on some finite set $\cU$,
such that $U\mkv A\mkv (B,E)$ form a Markov chain and
\begin{eqnarray*}
R 		&\geq& H(A|B) \ ,\\
\Delta	&\leq& \Big[ I(A;B|U) - I(A;E|U) \Big]_+ \ .
\end{eqnarray*}
\end{coro}

\subsection{Bob Has No Side Information}

\begin{coro}
\label{coro:noSI}
If Bob has no side information,
then region $\cR^*$ reduces to the set of all tuples $(R,D,\Delta)$ such that
there exist random variables $U$, $\hat A$ on $\cU$ and~$\cA$, respectively,
such that $U\mkv \hat A\mkv A\mkv E$ form a Markov chain and
\begin{eqnarray*}
R 		&\geq& I(\hat A;A) \ ,\\
D 		&\geq& \bE[d(A,\hat A)] \ ,\\
\Delta	&\leq& \Big[ H(A|\hat A) - I(A;E|U) \Big]_+ \ .
\end{eqnarray*}
\end{coro}

Note that, as opposed to the secure lossless source coding problem~\cite{gunduz2008secure},
in our situation, non-zero secrecy is still achievable when Bob has no side information.
In fact, Alice can exploit the available distortion at Bob to increase the equivocation rate at Eve.

\subsection{Bob Has Less Noisy Side Information Than Eve}

\begin{defi}
The side information $B$ is \emph{less noisy}
than the side information $E$ if
\[
I(U;B) \geq I(U;E) \ ,
\]
for each random variable $U$ such that
$U\mkv A\mkv (B,E)$ form a Markov chain.
\end{defi}

Note that the \emph{less noisy} condition is strictly weaker
than the \emph{stochastically degraded} one.

\begin{coro}
\label{coro:lessnoisy}
If Bob has less noisy side information than Eve, 
then region $\cR^*$ reduces to the set of all tuples $(R,D,\Delta)$ such that
there exist a random variable $V$ on some finite set $\cV$,
and a function $\hat A : \cV\times\cB \to \cA$
such that $V\mkv A\mkv (B,E)$ form a Markov chain and
\begin{eqnarray*}
R 		&\geq& I(V;A|B) \ ,\\
D 		&\geq& \bE[d(A,\hat A(V,B))] \ ,\\
\Delta	&\leq& \Big[ H(A|VB) + I(A;B) - I(A;E) \Big]_+ \ .
\end{eqnarray*}
\end{coro}

In this case, random variable~$U$ of Theorem~\ref{th:region} is set to a constant value
and hence Wyner-Ziv coding~\cite{wyner1976rate} achieves the optimal performance.

\subsection{Eve Has Less Noisy Side Information Than Bob}

\begin{coro}
If Eve has less noisy side information than Bob, 
then region $\cR^*$ reduces to the set of all tuples $(R,D,\Delta)$ such that
there exist a random variable $V$ on some finite set $\cV$,
and a function $\hat A : \cV\times\cB \to \cA$
such that $V\mkv A\mkv (B,E)$ form a Markov chain and
\begin{eqnarray*}
R 		&\geq& I(V;A|B) \ ,\\
D 		&\geq& \bE[d(A,\hat A(V,B))] \ ,\\
\Delta	&\leq& H(A|VE) \ .
\end{eqnarray*}
\end{coro}

In this case, random variable~$U$ of Theorem~\ref{th:region} is set to random variable $V$
and hence Wyner-Ziv coding~\cite{wyner1976rate} achieves the optimal performance.
Therefore it is not surprising that the equivocation rate at Eve corresponds to the case where Eve can reliably decode $V$.
Here, Alice can only exploit the available distortion at Bob to achieve a non-zero 
equivocation rate at Eve.

\section{Sketch of Proof of Theorem~\ref{th:region}}

\subsection{Proof of Achievability}
\label{sec:achievability}

In this section, we prove the achievability part of Theorem~\ref{th:region}
\emph{i.e.}, we prove the following proposition:

\begin{prop}
\label{prop:achievability}
Let $U$, $V$ be random variables on some finite sets $\cU$, $\cV$, respectively,
such that $U\mkv V\mkv A\mkv (B,E)$ form a Markov chain,
and $\hat A : \cV\times\cB \to \cA$.
If 
\begin{eqnarray*}
R 		&\geq& I(V;A|B) \ ,\\
D 		&\geq& \bE[d(A,\hat A(V,B))] \ ,\\
\Delta	&\leq& \Big[ H(A|VB) + I(A;B|U) - I(A;E|U) \Big]_+ \ ,
\end{eqnarray*}
then $(R,D,\Delta)$ is achievable.
\end{prop}

\emph{Proof:}
Let $\epsilon>0$ and define 
\begin{eqnarray*}
\Delta^* &=& \big[ H(A|VB) + I(A;B|U) - I(A;E|U) \big]_+	\ ,\\
\delta &=& \frac\epsilon{5 \cdot \max\left\{ d_{max}\ ,\ \Delta^* \right\}} \ .
\end{eqnarray*}

For a sufficiently large $n$, we build an $(n,R+\epsilon)$-code $(f,g)$
which achieves the required distortion and equivocation rate levels. 

\subsubsection{Codebook generation}

Randomly pick $2^{nS_1}$ sequences $u^n(s_1)$ from $\typ{U}$
and divide them into $2^{nR_1}$ equal size bins $\{B_1(r_1)\}_{r_1\in\{1,\dots,2^{nR_1}\}}$.
Then, for each codeword  $u^n(s_1)$,
randomly pick $2^{nS_2}$ sequences $v^n(s_1,s_2)$ from $\typ{V|u^n(s_1)}$
and divide them into $2^{nR_2}$ equal size bins $\{B_2(s_1,r_2)\}_{r_2\in\{1,\dots,2^{nR_2}\}}$.

\subsubsection{Encoding}

Assume that sequence $A^n$ is produced at Alice.
Look for a codeword $u^n(s_1)$ such that $(u^n(s_1),A^n)\in\typ{U,A}$.
Then look for a codeword $v^n(s_1,s_2)$ such that $(v^n(s_2),A^n)\in\typ{V,A|u^n(s_1)}$.
Let $B_1(r_1)$ and $B_2(s_1,r_2)$ be the bins of $u^n(s_1)$ and $v^n(s_1,s_2)$, respectively.
Alice sends the message $f(A^n) \triangleq (r_1,r_2)$ on the error-free channel.

\subsubsection{Decoding}

Assume that Bob receives $(r_1,r_2)$ from Alice and his side information sequence $B^n$.
Now look for the unique codeword $u^n(s_1)\in B_1(r_1)$ such that $(u^n(s_1),B^n)\in\typ{U,B}$.
Then look for the unique codeword $v^n(s_1,s_2)\in B_2(s_1,r_2)$ such that $(v^n(s_1,s_2),B^n)\in\typ{V,B|u^n(s_1)}$.
Compute the estimate $g(r_1,r_2,B^n)\in\cA^n$ using the component-wise
relation $g_i(r_1,r_2,B^n) \triangleq \hat A(v_i(s_1,s_2),B_i)$ for each $i=\{1,\dots,n\}$.

\subsubsection{Errors and constraints}

\begin{itemize}
\item For $n$ large enough, $\pr{(A^n,B^n)\not\in\typ{A,B}}<\delta$.
\item In the first encoding step, Alice needs to find (at least) one codeword 
$u^n(s_1)$ such that $(u^n(s_1),A^n)\in\typ{U,A}$.
If $S_1>I(U;A)$, then the probability that this step fails can be upper bounded
by $\delta$ for a sufficiently large $n$.
Similarly, the second encoding step requires the condition $S_2>I(V;A|U)$ to succeed
with a probability higher than $1-\delta$.
\item In the first decoding step, Bob looks for the \emph{unique} codeword $u^n(s_1)\in B_1(r_1)$ 
such that $(u^n(s_1),B^n)\in\typ{U,B}$. If $S_1-R_1<I(U;B)$, then the probability that
there exists another admissible codeword can be lowered below $\delta$ for a large $n$.
Similarly, the second decoding step requires the condition $S_2-R_2<I(V;B|U)$ to succeed
with a probability higher than $1-\delta$.
\end{itemize}

Defining the global transmitted rate $R+\epsilon=R_1+R_2$ and 
putting all inequalities together, we prove that
a sufficient condition for the above code to work with an error probability
lower than $5\delta$ is given by:
\begin{eqnarray*}
R+\epsilon
	&=& R_1+R_2 \\
	&>& I(U;A) - I(U;B) + I(V;A|U) - I(V;B|U) \\
	&\stackrel{(a)}{=}& I(V;A) - I(V;B) \\
	&\stackrel{(b)}{=}& I(V;A|B) \ ,
\end{eqnarray*}
where step~$(a)$, resp.~$(b)$, follows from the Markov chain $U\mkv V\mkv (A,B)$,
resp. $V\mkv A\mkv B$.
This condition is verified under the given assumption on $R$ \emph{i.e.}, $R \geq I(V;A|B)$.

\subsubsection{Distortion at Bob}

Denote by $F$ the event ``An error occurred during the encoding or decoding steps.''
We now check that our code achieves the required distortion level at Bob:
\begin{eqnarray*}
\bE\left[ d(A^n,g(f(A^n),B^n)) \right]
	&\leq& \pr{\bar{F}} \bE\left[ d(A^n,g(r_1,r_2,B^n)) \,\Big| \bar{F} \right]
				+ \pr{F}\,d_{max}\\
	&\leq& \dfrac1n \sum_{i=1}^n \bE\left[ d(A_i,\hat A(v_i(s_1,s_2),B_i)) \,\Big| \bar{F} \right] + \epsilon \\
	&=&	\bE\left[ d(A,\hat A(V,B)) \right] + \epsilon \\
	&\leq&	D + \epsilon \ .
\end{eqnarray*}

\subsubsection{Equivocation Rate at Eve}

The equivocation rate at Eve can be lower bounded as follows:
\begin{eqnarray*}
\frac1n\,H(A^n|f(A^n),E^n)
	&\geq&					\frac{\pr{\bar{F}}}n\,H(A^n|r_1 r_2 E^n, \bar{F}) \\
	&\geq& 					\frac{1-5\delta}n	\Big[ H(A^n) - I(A^n;r_1 E^n) - I(A^n;r_2|r_1 E^n)	\Big] \\
	&\geq&					\frac{1-5\delta}n	\Big[ H(A^n) - I(A^n;U^n E^n) - H(r_2) 				\Big] \\
	&=&						(1-5\delta) 		\Big[ H(A) - I(A;UE) - \frac1n\,H(r_2) 				\Big] \\
	&\geq&	(1-5\delta)			\Big[ H(A|UE) - R_2 								\Big] \\
	&\geq&	\Big[ H(A|UE) - R_2 \Big] - 5\delta \Big[ H(A|UE) - R_2 \Big] \ .
\end{eqnarray*}

If $R_2 > I(V;A|U) - I(V;B|U)$, 
the equivocation rate can thus be bounded as follows:
\begin{eqnarray*}
\frac1n\,H(A^n|f(A^n),E^n)
	&\geq& \Big[ H(A|UE) - R_2 \Big] - 5\delta \Big[ H(A|UE) - I(V;A|U) + I(V;B|U) \Big]_+ \\
	&\geq& \Big[ H(A|UE) - R_2 \Big] - \epsilon \ ,
\end{eqnarray*}
where the last inequality follows after some straightforward derivation
from the definition of~$\delta$ and the Markov chain $U\mkv V\mkv A\mkv B$.

Thus, each $\Delta \leq H(A|UE) - R_2$ is achievable. 
The above constraint on $R_2$ together with the latter inequality yield
the given sufficient condition.

\endproof

Note that our coding scheme can be interpreted as a simple binning operation to transmit~$U$,
followed by a Wyner--Ziv coding~\cite{wyner1976rate} to transmit~$A$ with side information $(U,B)$ at Bob.

\subsection{Proof of Converse}
\label{sec:converse}

In this section, we prove the converse part of Theorem~\ref{th:region}
\emph{i.e.}, we prove the following proposition:

\begin{prop}
\label{prop:converse}
Let $(R,D,\Delta)$ be an achievable tuple.
There exist two random variables $U$, $V$
and a function $\hat A : \cV\times\cB \to \cA$,
such that $U\mkv V\mkv A\mkv (B,E)$ form a Markov chain
and
\begin{eqnarray*}
R 		&\geq& I(V;A|B) \ ,\\
D 		&\geq& \bE[d(A,\hat A(V,B))] \ ,\\
\Delta	&\leq& \Big[ H(A|VB) + I(A;B|U) - I(A;E|U) \Big]_+ \ .
\end{eqnarray*}
\end{prop}

\emph{Proof:}
Let $\epsilon>0$.
There exists an $(n,R+\epsilon)$-code $(f,g)$ s.t.:
\begin{eqnarray*}
\bE\left[ d(A^n,g(f(A^n),B^n)) \right]	&\leq& D+\epsilon \ ,\\
\dfrac1n\,H(A^n|f(A^n),E^n) 			&\geq& \Delta-\epsilon \ .
\end{eqnarray*}

Denote by $W=f(A^n)$ the transmitted message.
The following Markov chain holds for each subset $J\subset\{1,\dots,n\}$:\footnote{
	$J^c$ denotes the complement of $J$ in $\{1,\dots,n\}$: $J^c = \{1,\dots,n\} \setminus J$.
}
\begin{equation}
\label{eq:chaine}
(W,A_J,B_J,E_J)\mkv A_{J^c} \mkv (B_{J^c},E_{J^c}) \ .
\end{equation}

For each $i\in\{1,\dots,n\}$, define $U_i$ and $V_i$
such that 
\begin{eqnarray}
U_i &=& (W,B_{i+1}^n,E^{i-1}) 					\label{eq:defU} \ ,\\
V_i &=& (W,A^{i-1},B^{i-1},B_{i+1}^n,E^{i-1})	\label{eq:defV} \ .
\end{eqnarray}

From Equation~\eqref{eq:chaine}, $U_i\mkv V_i\mkv A_i\mkv (B_i,E_i)$
form a Markov chain.

\subsubsection{Rate}

\begin{eqnarray*}
n(R+\epsilon)
	&\geq& 					H(W) \\
	&\stackrel{(a)}{=}&		I(W ; A^n B^n E^n) \\
	&\stackrel{(b)}{\geq}&	I(W ; A^n E^n | B^n) \\
	&\stackrel{(c)}{=}&		\sum_{i=1}^n I(W ; A_i E_i | A^{i-1} B^n E^{i-1}) \\
	&=&						\sum_{i=1}^n I(W A^{i-1} B^{i-1} B_{i+1}^n E^{i-1} ; A_i E_i | B_i) - I(A^{i-1} B^{i-1} B_{i+1}^n E^{i-1} ; A_i E_i | B_i) \\
	&\stackrel{(d)}{=}&		\sum_{i=1}^n I(W A^{i-1} B^{i-1} B_{i+1}^n E^{i-1} ; A_i E_i | B_i) \\
	&\stackrel{(e)}{\geq}&	\sum_{i=1}^n I(V_i ; A_i | B_i) \ ,
\end{eqnarray*}
where
\begin{itemize}
\item step~$(a)$ follows from $W=f(A^n)$,
\item step~$(b)$ from the non-negativity of mutual information,
\item step~$(c)$ from the chain rule for conditional mutual information,
\item step~$(d)$ from the independence of the random variables $A$, $B$ and $E$ across time,
\item step~$(e)$ from the non-negativity of mutual information and Equation~\eqref{eq:defV}.
\end{itemize}

Following the usual technique, we now define an independent random variable $Q$
uniformly distributed over the set $\{1,\dots,n\}$, and $A=A_Q$, $B=B_Q$, $E=E_Q$,
$U=(Q,U_Q)$, and $V=(Q,V_Q)$.
$U\mkv V\mkv A\mkv (B,E)$ still form a Markov chain and $(A,B,E)$ is distributed
according to the joint distribution $p(a,b,e)$ \emph{i.e.}, the original 
distribution of $(A_i,B_i,E_i)$. Then
\begin{eqnarray}
R + \epsilon
 	&\geq&	\frac1n \sum_{i=1}^n I(V_Q ; A_Q | B_Q, Q=i) \nonumber\\
	&=&		I(V_Q ; A_Q | B_Q Q) \nonumber\\
	&=&		I(V;A|B) \label{eq:converse_rate}\ .
\end{eqnarray}

\subsubsection{Distortion at Bob}

Bob reconstructs $g(W,B^n)$. 
The $i$-th coordinate of this estimate writes
\[
g_i(W,B^{i-1},B_i,B_{i+1}^n) \triangleq	\hat A_i(V_i,B_i) \ .
\]

The component-wise mean distortion at Bob is:
\begin{eqnarray*}
\bE\big[ d(A^n,g(f(A^n),B^n)) \big]
	&=& \frac1n \sum_{i=1}^n \bE[ d(A_i,\hat A_i(V_i,B_i)) ] \\
	&=& \frac1n \sum_{i=1}^n \bE[ d(A_Q,\hat A_Q(V_Q,B_Q))\ |\ Q=i] \\
	&=& \bE\big[ d(A_Q,\hat A_Q(V_Q,B_Q)) \big] \\
	&=& \bE\big[ d(A,\hat A(V,B)) \big] \ ,
\end{eqnarray*}
where we defined function $\hat A$ on $\cV\times\cB$ by 
\[
\hat A(V,B) = \hat A(Q,V_Q,B_Q) \triangleq \hat A_Q(V_Q,B_Q) \ .
\]

Consequently,
\begin{equation}
\label{eq:converse_distortion}
\bE[ d(A,\hat A(V,B)) ] \leq D + \epsilon \ .
\end{equation}

\subsubsection{Equivocation Rate at Eve}

\begin{eqnarray*}
H(A^n|W,E^n)
	&=& 					H(A^n|W) - I(A^n ; E^n | W) \\
	&=&					H(A^n|W B^n) + I(A^n;B^n | W) - I(A^n ; E^n | W) \\
	&\stackrel{(a)}{=}&	H(A^n|W B^n) + I(A^n ; B^n) - I(W ; B^n) - I(A^n ; E^n) + I(W ; E^n) \\
	&\stackrel{(b)}{=}&	\sum_{i=1}^n H(A_i|W A^{i-1} B^n) + I(A_i;B_i) - I(A_i ; E_i) - I(W B_{i+1}^n ; B_i) + I(W E^{i-1}; E_i) \\
	&\stackrel{(c)}{=}&	\sum_{i=1}^n H(A_i|W A^{i-1} B^n E^{i-1}) + I(A_i;B_i) - I(A_i ; E_i) - I(W B_{i+1}^n ; B_i) \\
	&&\hspace{2.7cm} +\ I(W E^{i-1}; E_i) + I(E_i ; B_{i+1}^n | W E^{i-1}) - I(B_i ; E^{i-1} | W B_{i+1}^n) \\
	&=&					\sum_{i=1}^n H(A_i|W A^{i-1} B^n E^{i-1}) + I(A_i;B_i) - I(A_i ; E_i) \\
	&&\hspace{5.5cm} +\ I(E_i ; W B_{i+1}^n E^{i-1}) - I(B_i ; W B_{i+1}^n E^{i-1})\\
	&\stackrel{(d)}{=}&	\sum_{i=1}^n H(A_i|V_i B_i) + I(A_i;B_i) - I(A_i ; E_i)	+ I(E_i ; U_i) - I(B_i ; U_i) \\
	&\stackrel{(e)}{=}&	\sum_{i=1}^n H(A_i|V_i B_i) + I(A_i;B_i|U_i) - I(A_i;E_i|U_i) \ ,
\end{eqnarray*}
where 
\begin{itemize}
\item step~$(a)$ follows from the Markov chain $W\mkv A^n\mkv (B^n,E^n)$,
\item step~$(b)$ from the chain rules for conditional entropy and mutual information,
	and the fact that random variables $A_i$, $B_i$ and $E_i$ are independent across time,
\item step~$(c)$ from the Markov chain $A_i\mkv W A^{i-1}\mkv E^{i-1}$ (see Equation~\eqref{eq:chaine})
	and Csiszar and Korner's equality~\cite{csiszar1978broadcast},
\item step~$(d)$ from the definitions of random variables $U_i$ and $V_i$ (Equations~\eqref{eq:defU} and~\eqref{eq:defV}, resp.),
\item step~$(e)$ from the Markov chain $U_i\mkv A_i\mkv (B_i,E_i)$.
\end{itemize}

Now, using auxiliary random variable $Q$ defined above,
\begin{eqnarray*}
\frac1n H(A^n|W,E^n)
	&=&	\frac1n \sum_{i=1}^n H(A_Q|V_Q B_Q, Q=i) + I(A_Q;B_Q|U_Q, Q=i) - I(A_Q;E_Q|U_Q, Q=i) \\
	&=&	H(A|V B) + I(A;B|U) - I(A;E|U) \ .
\end{eqnarray*}

Moreover, $H(A^n|W,E^n)\geq0$, consequently,
\begin{equation}
\label{eq:converse_equivocation}
\Big[  H(A|V B) + I(A;B|U) - I(A;E|U) \Big]_+ \geq \Delta - \epsilon \ .
\end{equation}

This proves Proposition~\ref{prop:converse}.

\endproof

\section{Application Example: Coding Binary Source with BEC and BSC Side Informations}

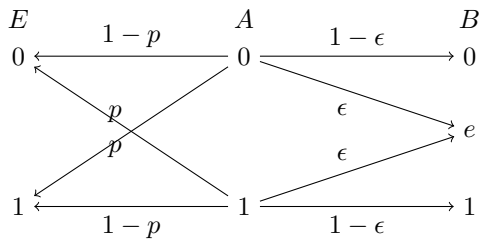
\begin{figure}[!b]
\centering
\begin{tikzpicture}
	\node	(A) 	at (0,.5) 	{$A$};
	\node	(A0) 	at (0,0) 	{$0$};
	\node	(A1)	at (0,-2)	{$1$};
	
	\node	(B) 	at (3,.5) 	{$B$};
	\node	(B0) 	at (3,0) 	{$0$};
	\node	(Be)	at (3,-1)	{$e$};
	\node	(B1)	at (3,-2)	{$1$};
	
	\node	(E) 	at (-3,.5) 	{$E$};
	\node	(E0) 	at (-3,0) 	{$0$};
	\node	(E1)	at (-3,-2)	{$1$};
	
	\draw[->] (A0) to node[auto,swap]{$1-p$}		(E0);
	\draw[->] (A0) to node[auto,swap]{$p$}			(E1);
	\draw[->] (A1) to node[auto]{$p$}				(E0);
	\draw[->] (A1) to node[auto]{$1-p$}				(E1);
	\draw[->] (A0) to node[auto]{$1-\epsilon$}		(B0);
	\draw[->] (A0) to node[auto,swap]{$\epsilon$}	(Be);
	\draw[->] (A1) to node[auto]{$\epsilon$}		(Be);
	\draw[->] (A1) to node[auto,swap]{$1-\epsilon$}	(B1);
\end{tikzpicture}
\caption{Considered model for source and side informations.}
\label{fig:example}
\end{figure}

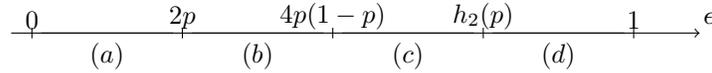
\begin{figure}[!h]
\centering
\begin{tikzpicture}
	\node	(ci)	at (-.4,0)	{};
	\node	(c0) 	at (0,0) 	{};
	\node	(c1)	at (2,0)	{};
	\node	(c2)	at (4,0)	{};
	\node	(c3)	at (6,0)	{};
	\node	(c4)	at (8,0)	{};
	\node	(e) 	at (9,0) 	{};
	
	\draw[->] (ci) to (e);
	\draw (c0) -- (c1) node[midway,anchor=north] {$(a)$};
	\draw (c1) -- (c2) node[midway,anchor=north] {$(b)$};
	\draw (c2) -- (c3) node[midway,anchor=north] {$(c)$};
	\draw (c3) -- (c4) node[midway,anchor=north] {$(d)$};
	\draw (0,2pt) -- (0,-2pt)	node[anchor=south]	{$0$};
	\draw (2,2pt) -- (2,-2pt)	node[anchor=south]	{$2p$};
	\draw (4,2pt) -- (4,-2pt) 	node[anchor=south] 	{$4p(1-p)$};
	\draw (6,2pt) -- (6,-2pt) 	node[anchor=south] 	{$h_2(p)$};
	\draw (8,2pt) -- (8,-2pt) 	node[anchor=south] 	{$1$};
	\draw (e) node[anchor=south] 	{$\epsilon$};
\end{tikzpicture}
\caption{The different regions as a function of $\epsilon$.}
\label{fig:cas}
\end{figure}

Consider the source model depicted in Fig.~\ref{fig:example}
where the source is binary and the side informations at Bob and Eve
are the outputs of a binary symmetric channel (BSC) with crossover probability $p\in[0,1/2]$ and a binary erasure channel (BEC) with erasure probability $\epsilon\in[0,1/2]$, respectively, with input $A$. Let $h_2$ denotes the binary entropy function given by $h_2(x) = -x\log_2(x) -(1-x)\log_2(1-x)$. According to the values of the parameters $(p,\epsilon)$ as summarized in Fig.~\ref{fig:cas}, it is not difficult to show by means of standard manipulations that the broadcast channel with input $A$ and outputs $(B,E)$ satisfies the following properties:
\begin{enumerate}
\item[(a)] The side information $E$ is a stochastically degraded version of $B$, \emph{i.e.}, 
	there exists a random variable $\tilde E$ such that $A\mkv B\mkv \tilde E$
	form a Markov chain and $P_{\tilde E|A}=P_{E|A}$,
\item[(b)] The side information $B$ is lessnoisy than $E$, \emph{i.e.},
	for all random variable $U$ such that $U\mkv A\mkv (B,E)$,
	$I(U;B)\geq I(U;E)$,
\item[(c)] The side information $B$ is more capable than $E$, \emph{i.e.},
	$I(A;B)\geq I(A;E)$,
\item[(d)] Any of the above relations hold between the side informations $B$ and $E$.
\end{enumerate}

Observe that this model is of interest since neither Bob nor Eve can always  be a lessnoisy decoder for all values of $(p,\epsilon)$. Thus in general $U$ is neither constant nor equal to $V$. We also remark that Corollary~\ref{coro:lessnoisy} provides the {rate-distortion-equivocation} region when $\epsilon$ lies in regions~$(a)$ or $(b)$. Otherwise, only Theorem~\ref{th:region} applies for the general case.

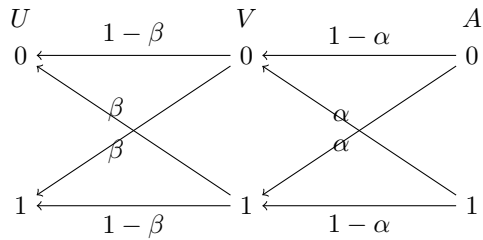
\begin{figure}[!b]
\centering
\begin{tikzpicture}
	\node	(A) 	at (0,.5) 	{$A$};
	\node	(A0) 	at (0,0) 	{$0$};
	\node	(A1)	at (0,-2)	{$1$};
		
	\node	(V) 	at (-3,.5) 	{$V$};
	\node	(V0) 	at (-3,0) 	{$0$};
	\node	(V1)	at (-3,-2)	{$1$};
	
	\node	(U) 	at (-6,.5) 	{$U$};
	\node	(U0) 	at (-6,0) 	{$0$};
	\node	(U1)	at (-6,-2)	{$1$};
	
	\draw[->] (A0) to node[auto,swap]	{$1-\alpha$}	(V0);
	\draw[->] (A0) to node[auto,swap]	{$\alpha$}		(V1);
	\draw[->] (A1) to node[auto]		{$\alpha$}		(V0);
	\draw[->] (A1) to node[auto]		{$1-\alpha$}	(V1);
	\draw[->] (V0) to node[auto,swap]	{$1-\beta$}		(U0);
	\draw[->] (V0) to node[auto,swap]	{$\beta$}		(U1);
	\draw[->] (V1) to node[auto]		{$\beta$}		(U0);
	\draw[->] (V1) to node[auto]		{$1-\beta$}		(U1);
\end{tikzpicture}
\caption{Binary auxiliary random variables.}
\label{fig:aux}
\end{figure}

From now on, let the distortion function at Bob $d$ be the Hamming distance and assume for simplicity that the source is uniform distributed, i.e., $\pr{A=0}=\pr{A=1}=1/2$. We know from the cardinality constraints given in Proposition~\ref{prop:card} that it suffices to consider sets $\cU$ and $\cV$  such that $\norm{\cU}\leq 4$ and $\norm{\cV}\leq 12$. Moreover, from the following proposition, we can restrict our attention to the auxiliary variables $(U,V)$ obtained as the outputs of a degraded binary symmetric broadcast channel with input $A$, as it is depicted in Fig.~\ref{fig:aux}. Notice that $V$ is identical to the auxiliary variable used  by Wyner and Ziv~\cite{wyner1976rate} for the rate-distortion function of a binary source in the case where there is no eavesdropper.

\begin{prop}
\label{prop:binary_rv}
In the case considered in this section, and depicted in Fig.~\ref{fig:example}, region $\cR^*$ is the set of all tuples $(R,D,\Delta)$ such that there exist $\alpha,\beta\in[0,1/2]$ satisfying
\begin{eqnarray*}
R 		&\geq& \varepsilon\,(1-h_2(\alpha)) 										\ ,\\
D 		&\geq& \varepsilon\,\alpha 													\ ,\\
\Delta	&\leq& \Big[ \varepsilon\,h_2(\alpha) + (1-\varepsilon)\,h_2(\alpha\star\beta)
									- h_2(p\star\alpha\star\beta) + h_2(p) \Big]_+ 	\ .
\end{eqnarray*}
\end{prop}

\emph{Proof:}
The achievability part of Proposition~\ref{prop:binary_rv} is a direct application of Theorem~\ref{th:region}: define auxiliary random variables $U$ and $V$ as depicted in Fig.~\ref{fig:aux}, and function $\hat A$ on $\cV$ by $\hat A(v) = v$. Expressions of Proposition~\ref{prop:binary_rv} follow after some straightforward derivations.

The converse part needs more arguments. The proof is omitted here and will be provided in an extended version of this paper.
\endproof

\begin{table*}[!b]
\caption{Some achievable tuples and corresponding auxiliary random variables.}
\label{tab:numeric}
\centering
\begin{tabular}{l||cc|cc}
							& Lossless secure source coding	& Slepian-Wolf 	& Lossy secure source coding	& Wyner-Ziv  \\
\hline
Rate $R$					& 0.469							& 0.469			& 0.375							& 0.375	\\
Distortion $D$				& 0								& 0				& 0.015							& 0.015	\\
Equivocation Rate $\Delta$	& 0.039							& 0				& 0.133							& 0.126	\\
\hline
$\alpha$					& 0								& 0				& 0.031							& 0.031	\\
$\beta$						& 0.078							& 0				& 0.050							& 0 
\end{tabular}
\end{table*}

\begin{figure}[!t]
\centering
\includegraphics[width=9cm]{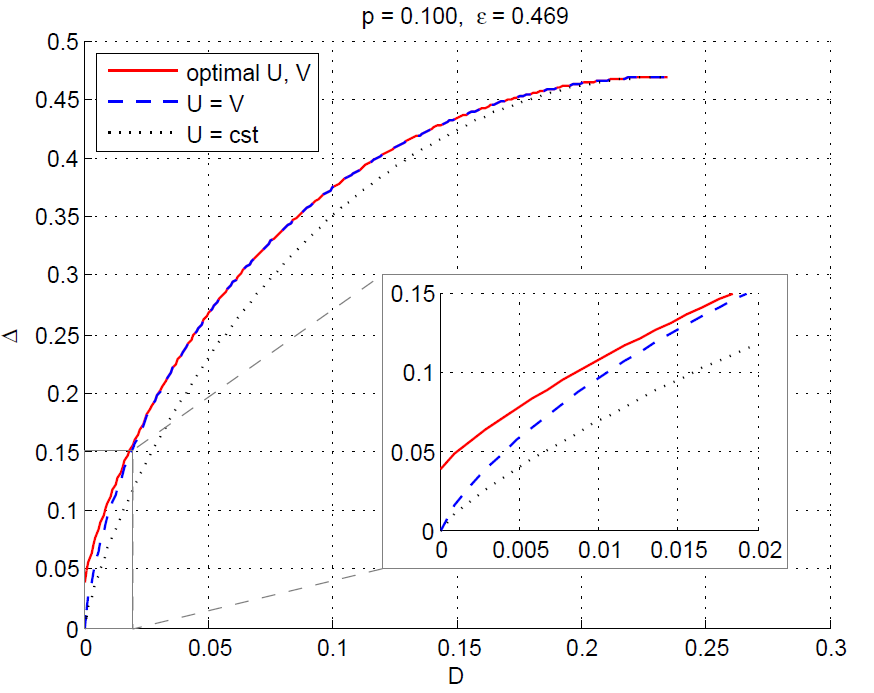}
\caption{Equivocation rate at Eve as a function of the distortion at Bob.}
\label{fig:Delta=fD}
\end{figure}

We now numerically compute some achievable values for $p=0.1$ and $\varepsilon=h_2(p)=0.469$ (see Fig.~\ref{fig:Delta=fD}).
In the case of lossless compression (columns \#1 and \#2 of Table~\ref{tab:numeric}),  the auxiliary random variable $V$ is set to be $A$ \emph{i.e.}, $\alpha=0$.
The additional variable $U$ actually enables a non-zero equivocation level,  as noted in~\cite{gunduz2008secure}. 
Assume that the coding rate is limited to a maximum of $80\%$ of the required rate for perfect reconstruction of the source (column \#3). This induces a distortion of $1.5\%$ at Bob and then an equivocation rate of $0.126\,$bits at Eve is achievable. This means that even a small increase in the distortion at Bob can be fully exploited by Alice to achieve very significant gains (more than third times in this case) in terms of equivocation rate at Eve.
Moreover, in the situation considered in this paragraph, Wyner-Ziv coding actually achieves the optimal performance for distortion levels higher than $0.036$ as shown in Fig.~\ref{fig:Delta=fD}

\section{Summary and Discussions}
The problem of secure lossy source coding of memoryless sources in the presence of an eavesdropper with different correlated side informations at the legitimate decoder (Bob) and the eavesdropper (Eve) was investigated. A complete characterization of the rate-distortion-equivocation region was derived for the case of arbitrary correlated side information at the decoders. It was shown that both the available distortion at the legitimate decoder and the statistical properties of the side informations can be exploited by the encoder (Alice) to increase the equivocation rate at the eavesdropper.

An application example to secure lossy source coding of a binary source, where the side information at Bob (resp. Eve) is the output of a binary erasure channel (resp. a binary symmetric channel) with the source as the input, was considered. This model is of interest since neither Bob nor Eve can always be a lessnoisy decoder and thus the encoding strategy needed to achieve the optimal equivocation rate is rather tricky. 
In the case where the source is uniformly distributed, we proved that it is optimal to consider binary auxiliary random variables and provided corresponding expressions for the rate-distortion-equivocation region.

As future and on-going work, it would be of interest to extend the results in the present work to the more general setting in which the encoder wishes to maximize not only the equivocation rate, but also an arbitrary measure of the equivocation distortion at the eavesdropper.

\bibliographystyle{IEEEtran}
\bibliography{allerton}

\begin{thebibliography}{10}
\providecommand{\url}[1]{#1}
\csname url@samestyle\endcsname
\providecommand{\newblock}{\relax}
\providecommand{\bibinfo}[2]{#2}
\providecommand{\BIBentrySTDinterwordspacing}{\spaceskip=0pt\relax}
\providecommand{\BIBentryALTinterwordstretchfactor}{4}
\providecommand{\BIBentryALTinterwordspacing}{\spaceskip=\fontdimen2\font plus
\BIBentryALTinterwordstretchfactor\fontdimen3\font minus
  \fontdimen4\font\relax}
\providecommand{\BIBforeignlanguage}[2]{{%
\expandafter\ifx\csname l@#1\endcsname\relax
\typeout{** WARNING: IEEEtran.bst: No hyphenation pattern has been}%
\typeout{** loaded for the language `#1'. Using the pattern for}%
\typeout{** the default language instead.}%
\else
\language=\csname l@#1\endcsname
\fi
#2}}
\providecommand{\BIBdecl}{\relax}
\BIBdecl

\bibitem{slepian1973noiseless}
D.~Slepian and J.~Wolf, ``Noiseless coding of correlated information sources,''
  \emph{IEEE Trans. Inf. Theory}, vol.~19, no.~4, pp. 471--480, 1973.

\bibitem{wyner1976rate}
A.~Wyner and J.~Ziv, ``The rate-distortion function for source coding with side
  information at the decoder,'' \emph{IEEE Trans. Inf. Theory}, vol.~22, no.~1,
  pp. 1--10, 1976.

\bibitem{shannon1949communication}
C.~Shannon, ``Communication theory of secrecy systems,'' \emph{BSTJ}, vol.~28,
  pp. 656--715, 1949.

\bibitem{wyner1975wire}
A.~Wyner, ``The wire-tap channel,'' \emph{BSTJ}, vol.~54, no.~8, pp.
  1355--1387, 1975.

\bibitem{csiszar1978broadcast}
I.~Csiszar and J.~Korner, ``Broadcast channels with confidential messages,''
  \emph{IEEE Trans. Inf. Theory}, vol.~24, no.~3, pp. 339--348, 1978.

\bibitem{it2008special}
``Special issue on information theoretic security,'' \emph{IEEE Trans. Inf.
  Theory}, vol.~54, no.~6, pp. 2405--2818, 2008.

\bibitem{liang2009information}
Y.~Liang, H.~Poor, and S.~Shamai, \emph{Information theoretic security}.\hskip
  1em plus 0.5em minus 0.4em\relax Now Publishers, 2009.

\bibitem{yamamoto1983source}
H.~Yamamoto, ``A source coding problem for sources with additional outputs to
  keep secret from the receiver or wiretappers,'' \emph{IEEE Trans. Inf.
  Theory}, vol.~29, no.~6, pp. 918--923, 1983.

\bibitem{yamamoto1988rate}
------, ``A rate-distortion problem for a communication system with a secondary
  decoder to be hindered,'' \emph{IEEE Trans. Inf. Theory}, vol.~34, no.~4, pp.
  835--842, 1988.

\bibitem{yamamoto1994coding}
------, ``Coding theorems for {Shannon's} cipher system with correlated source
  outputs, and common information,'' \emph{IEEE Trans. Inf. Theory}, vol.~40,
  no.~1, pp. 85--95, 1994.

\bibitem{yamamoto1997rate}
------, ``Rate-distortion theory for the {Shannon} cipher system,'' \emph{IEEE
  Trans. Inf. Theory}, vol.~43, no.~3, pp. 827--835, 1997.

\bibitem{liu2009securing}
R.~Liu and W.~Trappe, \emph{Securing wireless communications at the physical
  layer}.\hskip 1em plus 0.5em minus 0.4em\relax Springer, 2010.

\bibitem{merhav2006shannon}
N.~Merhav, ``On the {Shannon} cipher system with a capacity-limited
  key-distribution channel,'' \emph{IEEE Trans. Inf. Theory}, vol.~52, no.~3,
  pp. 1269--1273, 2006.

\bibitem{prabhakaran2007secure}
V.~Prabhakaran and K.~Ramchandran, ``On secure distributed source coding,'' in
  \emph{Proc. ITW}, 2007, pp. 442--447.

\bibitem{gunduz2008secure}
D.~Gunduz, E.~Erkip, and H.~Poor, ``Secure lossless compression with side
  information,'' in \emph{Proc. ITW}, 2008, pp. 169--173.

\bibitem{gunduz2008lossless}
------, ``Lossless compression with security constraints,'' in \emph{Proc.
  ISIT}, 2008, pp. 111--115.

\bibitem{tandon2009securea}
R.~Tandon, S.~Ulukus, and K.~Ramchandran, ``Secure source coding with a
  helper,'' in \emph{Proc. Allerton}, 2009, pp. 1061--1068.

\bibitem{merhav2008shannon}
N.~Merhav, ``{Shannon's} secrecy system with informed receivers and its
  application to systematic coding for wiretapped channels,'' \emph{IEEE Trans.
  Inf. Theory}, vol.~54, no.~6, pp. 2723--2734, 2008.

\bibitem{cover2006elements}
T.~Cover and J.~Thomas, \emph{Elements of information theory (2nd Ed)}.\hskip
  1em plus 0.5em minus 0.4em\relax Wiley-Interscience, 2006.

\end{thebibliography}

\end{document}